\documentclass{ws-ijmpd}
\begin{document}

%

\def\nocropmarks{\vskip5pt\phantom{cropmarks}}

 \let\trimmarks\nocropmarks      

%


\def\s{\mbox{\boldmath$\displaystyle\mathbf{\sigma}$}}
\def\J{\mbox{\boldmath$\displaystyle\mathbf{J}$}}
\def\K{\mbox{\boldmath$\displaystyle\mathbf{K}$}}

\def\beq{\begin{eqnarray}}
\def\eeq{\end{eqnarray}}


\def\cdash{$^{\raisebox{-0.5pt}{\hbox{--}}}$}         



\markboth{D. V. Ahluwalia-Khalilova, Irina Dymnikova}
{A theoretical case for negative mass-square for sub-eV particles}

%
\catchline{}{}{}
%

\title{A THEORETICAL CASE FOR NEGATIVE 
MASS-SQUARE \\FOR SUB-eV PARTICLES\footnote{This essay received an 
``honorable mention'' in the 2003 Essay Competition of the Gravity Research 
Foundation.}}


\author{\footnotesize D. V. AHLUWALIA-KHALILOVA\footnote{E-mail: 
d.v.ahluwalia-khalilova@heritage.reduaz.mx}}

\address{Center for Studies of Physical, Mathematical, and
Biological Structure of Universe\\
Department of Mathematics, Ap. Postal C-600\\
University of Zacatecas (UAZ), Zacatecas, ZAC 98062, Mexico}


\author{IRINA DYMNIKOVA\footnote{E-mail: irina@matman.uwm.edu.pl }}

\address{Department of Mathematics and 
Computer Science, University of Warmia and Mazury\\
\.Zo{\l}nierska 14, Olsztyn 10-561, Poland}


\maketitle

\begin{abstract}
Electroweak gauge bosons have masses of the order of 
$10^2\,\,\mbox{GeV} [/c^2]$, while masses of additional bosons
involved in gravito-electroweak unification are expected to be still
higher. These are  at least eleven orders of magnitude higher than
sub-eV range indications for neutrino masses. Under these
circumstances we suspect that the sub-eV particles are created in
a spacetime where gravitational effects of massive gauge bosons 
may become important.  
The question that we thus ask is: What is the spacetime
group around a gravito-electroweak vertex? Modeling it as de-Sitter
we find that sub-eV particles may carry
a negative mass square of the order of $- \left(3/8 \pi^3\right) \left(
M_{\mathrm{unif.}}/M_{\mathrm{Planck}}\right)^4 M_{\mathrm{Planck}}^2$.
Neutrino oscillation data then hints at $30-75$ TeV scale for 
$M_{\mathrm{unif.}}$, where 
$M_{\mathrm{unif.}}$ characterizes gravito-electroweak
unification scale.   
\end{abstract}


\section{Introduction}

One of the lessons from E. P. Wigner's early works\cite{EPW1939} is  
that the notion of mass is not an arbitrary physical construct but
takes its origin from constancy of speed of light for all inertial
observers. The latter 
implies description of
physical states in terms of the  Casimir invariants associated
with the Poincar\'e group:
\beq 
C_1= P_\mu P^\mu\,,\quad C_2 = W_\mu W^\mu\,,
\eeq
with Pauli-Lubanski pseudovector, $W_\mu$, defined as
\beq
W_\mu = -\frac{1}{2} \epsilon_{\mu\nu\rho\sigma} J^{\nu\rho} P^{\sigma}\,.
\eeq
Here we use the notation of Ref. [\refcite{LHR1996}].
Each representation space is then characterized  by eigenvalues 
of these Casimir operators. 
Representation spaces of the type $(j,0)\oplus(0,j)$ are  
characterized by a 
\begin{enumerate}
\item[(a)]
Positive definite mass, and
\item[(b)]
Single spin-$j$
\end{enumerate}
while 
spaces of the type $\left[(j,0)\oplus(0,j)\right](1/2,1/2)$
carry interpretation of \cite{ak2001,ka2002}
\begin{enumerate}
\item[(i)]
Positive definite mass, but
\item[(ii)]
Indefinite/multiple spin.
\end{enumerate}
Attempts to force a single-spin 
interpretation ~\textendash~
 as for Rarita-Schwinger 
framework ~\textendash~ result in well-known problems \cite{kl}. 
Efforts to implement a single-spin interpretation 
on  $\left[(j,0)\oplus(0,j)\right](1/2,1/2)$ spaces
 are akin to insisting on a
``particle'' interpretation for the Dirac's $(1/2,0)\oplus(0,1/2)$
representation space by imposing a covariant constraint which
throws away the ``negative energy'' sector, i.e., the antiparticles.
Or, at least
this is a view we have put forward in Refs. [\refcite{ak2001,ka2002}].

Now, with the discovery of massive gauge bosons of electroweak 
interactions a new situation has arisen. 
This, as we shall now argue, may question positive/real definiteness of 
mass for sub-eV particles.

The massive gauge bosons  have masses of the order of 
$10^2\,\,\mbox{GeV} [/c^2]$. These are at least 
eleven orders of magnitude higher than
sub-eV range indication for neutrino 
masses\cite{Davis1}\cdash\cite{LSND2}.
Our thesis arises from the possibility 
that the sub-eV particles are created in
a spacetime where gravitational effects of massive gauge bosons 
may become important.  

The question that we thus ask is: What is the spacetime
group around a gravito-electroweak vertex? 
In the context of Refs. [\refcite{id2002,id2003}],
if we impose the requirements of (a) spherical symmetry,
(b) dominant energy condition for a source term,
(c)  regularity of density, and (d) finiteness of mass;
then, the answer is de Sitter-Schwarzschild
geometry. Thus, in the interaction region
the spacetime symmetry group is de Sitter.
This may be taken as our fundamental working 
assumption.\footnote{This approximation  neglects 
spin of the massive gauge bosons which may, when
properly accounted for, may be responsible for explaining 
neutrino mixing matrix.}

For the  de Sitter-Schwarzschild case,
the stress-energy tensor evolves smoothly
from  de Sitter vacuum $T_{\mu\nu} = \rho_0 c^2 g_{\mu\nu}$
at $r=0$ to Minkowski vacuum $T_{\mu\nu}=0$ at infinity.
Here $\rho_0$ is the mass density at the origin, and shall be
identified with the gravito-electroweak scale.
The induced metric is given by\cite{id1992}:
\beq
ds^2=\left(1-\frac{2 G M(r)}{c^2 r}\right)c^2 dt^2 -
      \left(1-\frac{ 2 G M(r)}{c^2 r}\right)^{-1} dr^2 
       - r^2 d\theta^2 - r^2 \sin^2\theta \,d\phi^2\,\,,&&\nonumber\\
\eeq
with 
\beq
M(r) = 4 \pi\int_0^r \rho(x) x^2 dx\,.
\eeq
For any density profile satisfying the requirements
(b)-(d) enumerated above, asymptotic behavior in the
$r\rightarrow 0$ region is dictated by (b) and is de Sitter
vacuum:
\beq
ds^2= \left(1-\frac{r^2}{r_0^2}\right) c^2 dt^2 -
      \left(1-\frac{r^2}{r_0^2}\right)^{-1} dr^2 
       - r^2 d\theta^2 - r^2 \sin^2\theta \,d\phi^2\,\,,
\eeq  
with 
\beq
r_0^2=\frac{3 c^2}{8\pi G \rho_0}\,.\label{r0}
\eeq
For $r\rightarrow \infty$, the asymptotic is Schwarzschild
\beq
ds^2=
\left(1-\frac{2 G m}{c^2 r}\right) c^2 dt^2 -
      \left(1-\frac{ 2 G m}{c^2 r}\right)^{-1} dr^2 
       - r^2 d\theta^2 - r^2 \sin^2\theta \,d\phi^2\,,
\eeq
where $m= M(r\rightarrow \infty)$ is the ADM mass.

\section{Negative mass-square for sub-eV particles}

We now envisage that a sub-eV particle, which we may think of as neutrino
to be concrete,  
is created in the de Sitter region and 
later propagates to $r\rightarrow\infty$ (i.e., to r $\gg r_0$ region)
where spacetime is Minkowskian to a good approximation.
In the creation region, such a particle is characterized as an  
eigenstate of the de Sitter Casimir invariants, 
$\vert I_1^\prime, \;I_2^\prime\rangle$. The $I_1^\prime, \;I_2^\prime$ are
eigenvalues, respectively, of the de Sitter Casimir 
operators\cite{fg}:\footnote{Note that $\eta_{\mu\nu}$ of Ref. 
[\refcite{fg}] and that used here differ by 
a minus sign.}
\beq
&&I_1 =  \Pi_\mu \Pi^\mu - \frac{1}{2 r_0^2} J_{\mu\nu} J^{\mu\nu}\,,
\eeq 
and  $I_2$ (see, Ref. [\refcite{fg}], for its definition). 
In going from Poincar\'e  to de Sitter symmetries, the 
notion of mass must undergo an unavoidable change.
To investigate this modification, 
we shall concentrate on $I_1$ only.
The $\Pi_\mu$ is defined as,
\beq
\Pi_\mu = \left(  1+ \frac{r^2-c^2 t^2}{4 r_0^2}\right)P_\mu 
+ \frac{1}{2 r_0^2} {x^\nu} J_{\mu\nu}\,. 
\eeq 
In the interaction region $r^2-c^2 t^2 \ll r_0^2$.\footnote{It shall
 be confirmed explicitly below towards the end of next section.} So,
$I_1$  approximates to:
\beq
I_1 \approx  P_\mu P^\mu - \frac{1}{r_0^2}  \left(\J^2-\K^2\right)\,.
\eeq 
Or, equivalently
\beq
I_1 \approx C_1 - \frac{1}{r_0^2}  \left(\J^2-\K^2\right)\,,
\eeq
where, we remind,  $C_1$ is the first Casimir operator of the Poincar\'e group.
Eigenvalue of $C_1$, 
up to a multiplicative factor of $c^2$, is identified with
square of the mass of the particle, $\mu^2$.
 
In order to study implications for sub-eV neutrinos, 
we now evaluate the dragged $I_1$ 
for $(1/2,0)$ and $(0,1/2)$ representations spaces.\footnote{
See, Ref. [\refcite{ka2002}], for the definition
of a dragged Casimir. Also, 
recall that, 
$
J_{ij}=-J_{ji}= \epsilon_{ijk} J_k$ 
and 
$J_{i 0}=-J_{0 i} = - K_i$, 
with each of  the $i,j,k$ taking the values
$1,2,3$. The $\J$ are then generators of Lorentz 
rotations and $\K$ are
generators of Lorentz boosts.} 
The right handed and 
left handed fields inhabit these spaces, respectively.
For the $(1/2,0)$ representation space, we have \cite{LHR1996},
\beq
{\J} = \hbar \frac{\s}{2}\,,\quad
{\K} = -i \hbar \frac{\s}{2} \,,
\eeq
while for the $(0,1/2)$ representation space,
\beq
{\J} = \hbar \frac{\s}{2}\,,\quad
{\K} = +i \hbar \frac{\s}{2} \,.
\eeq
This, immediately yields:
\beq
I_1 \approx  P_\mu P^\mu  - \frac{\hbar^2}{2 r_0^2}\s^2\,.
\eeq
Its eigenvalues are:\footnote{~~Note, it corrects Ref. [\refcite{v1}].} 
\beq 
I_1^\prime = \mu^2 c^2 - \frac{3 \hbar^2}{2 r_0^2}\,.
\eeq
It is now explicit that 
the notion of mass is modified in going form one spacetime symmetry
group to another. 
As we shall shortly see, 
this  modification allows for negative mas-square for sub-eV particles
if gravito-electroweak unification occurs at TeV scales.

This is the central result of this essay and 
may offer a natural  explanation for
certain anomalous results which have 
come to be known  as ``negative mass squared problem'' 
for $\overline{\nu}_e$\cite{nms1}\cdash\cite{nms5},
and $\nu_\mu$\cite{nms6}\cdash\cite{nms8} even though,
as outlined in the Addenda, efforts in
data analysis tend toward imposing by hand the requirement
of physical $m_\nu^2 > 0$.

Thus the negative mass-squared values for sub-eV particles,
and  neutrinos in particular, 
may be expected to be governed by parameter:

\beq
m_{\mathrm {neg.}}^2 = -\,\frac{3 \hbar^2}{2 r_0^2 c^2}\,.
\eeq

\section{Hint for a TeV scale gravito-electroweak unification}

If $\rho_0$ is identified with an (yet unknown) electroweak-gravitation 
mass scale $M_{\mathrm{unif.}}$ then, on recalling the 
definition of $r_0$ from Eq. (\ref{r0}),
we have
\beq
m_{neg.}^2 &=& - \frac{4 \pi G \hbar^2}{c^4} \rho_0\,,\\
&=&  - \frac{4 \pi G \hbar^2}{c^4}\left( 
{M_{\mathrm{unif.}}}\bigg/\bigg[  \left( \frac{4\pi}{3} \right) 
\left( \frac{2\pi\hbar} {M_{\mathrm {unif.}} c} \right)^3  \bigg]\right)\,,\\           
&=& - \frac{3}{8 \pi^3} \frac{G}{\hbar c} M_{\mathrm {unif.}}^4
\eeq
Identifying, $\sqrt{{\hbar c}/{G}}$ with
$M_{\mathrm {Planck}}  $
 the above expression becomes,
\beq
m_{\mathrm{neg.}}^2
=-  \frac{3}{8 \pi^3}  \left( \frac{M_{\mathrm {unif.}}}{M_{\mathrm {Planck}}}
\right)^4 M^2_{\mathrm Planck}
\eeq
If $M_{\mathrm{unif.}}$ is set to be $100$ GeV, i.e. of the order of
masses for electroweak gauge bosons $W^\pm$ and $Z$,  
one immediately sees that  $m_{neg.}^2 
\simeq - 8.4 \times 10^{-15} \mbox{eV}^2$. Existing data on neutrino masses 
rules out this identification because it is natural to expect that 
$m_{neg.}^2 \sim - \Delta m^2$. Where, $\Delta m^2$ 
as derived from atmospheric and solar neutrino data is \cite{pv}:
$
\Delta m^2_{\sc ATM} = 2.5 \times 10^{-3} \,\mbox{eV}^2 [/c^4],\,\,
\Delta m^2_{\sc SOL} = 6.9 \times 10^{-5} \,\mbox{eV}^2 [/c^4]
$.   

However, 
$\left(M_{\mathrm {unif.}}/{M_{\mathrm {Planck}}}\right)^4$-sensitivity of $m_{\mathrm{neg.}}^2$ suggests 
a TeV scale for $M_{\mathrm{unif.}}$. This can be seen explicitly by 
setting $m_{neg.}^2 \sim - \Delta m^2$.  Then 
$M_{\mathrm unif.}$ reads:
\beq
{M_{\mathrm {unif.}}} \sim \left(
\frac{8 \pi^3}{3}
\frac{\Delta m^2}
     {M_{\mathrm {Planck}}^2}
                                  \right)^{1/4} 
{M_{\mathrm {Planck}}}
\eeq

The atmospheric neutrino data implies a $M_{unif.} \simeq 74$ TeV
while the cited solar neutrino mass-squared difference yields,
$M_{unif.} \simeq 30$ TeV. These $M_{unif.}$,
correspond, respectively,  to  $r_0\simeq 5\times 10^{-4}$ cm, and
$r_0 \simeq 3\times 10^{-3}$ cm.

\section{Conclusion}

In view of these considerations, and additional and earlier
work of  Simicevic\cite{ns}, it appears that to discard
experimentally indicated $m_\nu^2 < 0$ for electron and 
muon neutrinos may be unwise. The best route may be to
look at the data and experiments afresh and allow that 
it may indeed be that 
\beq
m_{\nu_e}^2 < 0\,,\quad m_{\nu_\mu}^2 <0\,.\label{oneagain}
\eeq
At the same time a global analysis of data on neutrinos
~\textendash~
specifically data on neutrino oscillations, data on neutrino-less
double beta decay, data on the end point of tritium beta decay,
and $\pi^+ \rightarrow \mu^+ + \nu_\mu$ ~\textendash~
must be done to allow for  $m_\nu^2 < 0$. If negative mass-square 
is finally established for a sub-eV particle it would be necessary
to device experiments which may distinguish between various 
proposals\cite{nms2}\cdash\cite{nms14} which suggest, or
attempt to accommodate, negative mass squares.

\section*{Addenda}
We think that   the following additional information
may be helpful to the reader of our essay:

\vspace*{12pt}
\noindent
\underline{(Anti)Electron neutrino mass}

\vspace*{7pt}
\noindent
 The latest publication on $m_{\nu_e}^2$ at 
the time of sending
this essay to IJMPD seems to be from Lobashev \cite{vmb}. 
It gives $m_{\mu_e} ^2 = - 2.5  \pm 2.5 \pm 2.0  \,\,\mbox{eV}^2$. 
Particle data group\cite{pdg2002}, gives: 
$m_{\nu_e} ^2 = - 2.5  \pm 3.3 \,\,\mbox{eV}^2$ and only includes results
of Lobashev\cite{L1999} and Weinheimer\cite{W1999} with the following two
observations: 
\begin{enumerate}
\item
The data were corrected for electron trapping effects in the source, eliminating the dependence of the fitted neutrino mass on the fit interval. 
The analysis assuming a pure beta spectrum yields 
significantly negative fitted $m_\nu^2 \approx - (20 -10)
\,\,\mbox{eV}^2$. This problem is attributed to a 
discrete spectral anomaly of about $6 \times 10^{- 11}$ intensity
 with a time-dependent energy of $5- 15\,\,\mbox{eV}$ 
below the endpoint. 
The data analysis accounts for this anomaly by introducing 
two extra phenomenological fit parameters resulting 
in a best fit of $m_\nu^2 =- 1.9\pm 3.4 \pm 2.2\,\, \mbox{eV}^2$ 
which is used 
to derive a neutrino mass limit. However, the introduction 
of phenomenological fit parameters which are correlated 
with the derived $m_\nu^2$  limit makes unambiguous interpretation 
of this result difficult. 
\item
We do not use the following data for averages, fits, limits, etc.
$\ldots$ Stoeffl\cite{nms1}:  $m_\nu^2 = -130\pm 20\pm 15\,\,\mbox{eV}^2$ 
$\ldots$ Robertson\cite{nms9}:  $m_\nu^2 = -147\pm 68\pm 41\,\,
\mbox{eV}^2$.\footnote{
Taking  $m^2_{\mathrm neg.}\sim -10^2\,\, \mbox{eV}^2$ yields
$M_{\mathrm{unif}}  \sim 10^3\,\,\mbox{TeV}$.}

\end{enumerate}
In order to put all these matters in perspective one of us has taken 
liberty of asking  one of the early experimentalists about his
views on his own experiment. His candid reply reads,
\begin{quote} 
We still have no clue why there was an excess of counts close to the 
beta-endpoint in our spectra. The experiment is ``mothballed'' since 1993 and 
has not been operated since.  We were not able to come up with a 
satisfactory explanation for the bump at the end of the spectrum. In the 
meantime, The neutrino group in $\ldots$ improved their 
spectrometers. They had initially similar puzzling result, but in the 
meantime, the neutrino mass extracted from their spectra is consistent with 
zero or a very small value.
\end{quote}
and further strengthens the need for careful experiments without any prejudice
in data analysis for $m^2_\nu> 0$.

\vspace*{12pt}
\noindent
\underline{Muon neutrino mass}

\vspace*{7pt}
\noindent
Additional hint for negative mass-square for neutrino masses
resides in two possible values ~\textendash ~ determined by which value of pion
mass one uses ~\textendash ~ 
for $m_{\nu_\mu}^2$:\cite{pdg2002}\footnote{Taking  
$m^2_{\mathrm neg.}\simeq - 0.143\,\, \mbox{MeV}^2$ yields
$M_{\mathrm{unif}}  \simeq 2.0 \times 10^5\,\,\mbox{TeV}$.}

\beq
&&\mbox{Solution B:} \quad- 0.016 \pm 0.023\,\, 
\mbox{MeV}^2\,,\nonumber \\
&&\mbox{Solution A:} \quad -0.143 \pm 0.024 \,\, \mbox{MeV}^2\,.\nonumber
\eeq

\section*{Acknowledgments}

DVA-K thanks colleagues in neutrino community for clarification on the 
status of experiments on beta-endpoint spectra and Majorana/Dirac 
nature of neutrinos,  
Irina Dymnikova for her warm hospitality  in Olsztyn,   a
referee for his helpful remarks, and  N. Simicevic for bringing his
work on the subject to our attention. This published version
was duly updated after we learned of Simicevic's work. 

This work was supported by the Polish Committee for 
Scientific Research through the grant 5P03D.007.20. Additional support 
was provided  by CONACyT Project 32067-E.


\end{document}